\documentclass[aps,prl,reprint,groupedaddress]{revtex4-1}

\usepackage{amsmath}
\usepackage{amscd}
\usepackage[xdvi]{graphicx}
\usepackage{txfonts}

\begin{document}

\title{Center manifold reduction for large populations of globally coupled phase oscillators}

\author{Hayato Chiba}
\affiliation{Faculty of Mathematics, Kyushu University, Fukuoka 819-0395, Japan}

\author{Isao Nishikawa}
\affiliation{Department of Mathematical Informatics,
Graduate School of Information Science and Technology, University of Tokyo, Tokyo 113-8656, Japan}

\date{\today}

\begin{abstract}
A bifurcation theory for a system of globally coupled phase oscillators is developed based on the 
theory of rigged Hilbert spaces. It is shown that there exists a finite-dimensional center manifold
on a space of generalized functions. 
The dynamics on the manifold is derived for any coupling functions. 
When the coupling function is $\sin \theta $, a bifurcation diagram conjectured by Kuramoto is rigorously obtained. 
When it is not $\sin \theta $, a new type of bifurcation phenomenon is 
found due to the discontinuity of the projection operator to the center subspace.
\end{abstract}


\maketitle







\begin{figure}
\includegraphics{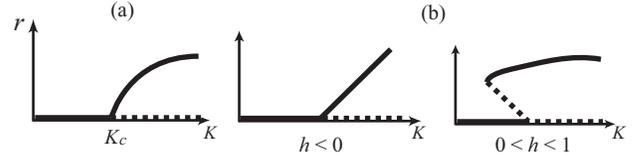}
\caption{Bifurcation diagrams of the order parameter for (a) $f(\theta ) = \sin \theta $
and (b) $f(\theta ) = \sin \theta  + h\sin 2\theta$.
The solid lines denote stable solutions, and the dotted lines denote unstable solutions. \label{fig1}}
\end{figure}

\textbf{Introduction}.
Collective synchronization phenomena are observed in a variety of areas, such as chemical reactions,
engineering circuits, and biological populations~\cite{PRK}.
In order to investigate such phenomena, a system of globally coupled phase oscillators of the following form is often used:
\begin{equation}
\frac{d\theta _i}{dt} 
= \omega _i + \frac{K}{N} \sum^N_{j=1} f(\theta _j - \theta _i),\,\, i= 1, \cdots  ,N,
\label{KM}
\end{equation}
where $\theta _i(t) $ denotes the phase of an $i$-th oscillator,
$\omega _i\in \mathbf{R}$ denotes its natural frequency drawn from some distribution function $g(\omega )$, 
$K>0$ is the coupling strength, and $f(\theta ) = \sum^\infty_{n=-\infty}f_n e^{\sqrt{-1}n\theta }$ is a $2\pi$-periodic function.
When $f(\theta ) = \sin \theta $, it is referred to as the Kuramoto model~\cite{Kura2}.
In this case, it is numerically observed that if $K$ is sufficiently large,
some or all of the oscillators tend to rotate at the same velocity on average, which is referred to as 
\textit{synchronization}~\cite{PRK,Str1}.
In order to evaluate whether synchronization occurs, Kuramoto introduced
the order parameter $r(t)e^{\sqrt{-1}\psi (t)}$, which is given by
\begin{equation}
r(t)e^{\sqrt{-1}\psi (t)} := \frac{1}{N}\sum^N_{j=1} e^{\sqrt{-1} \theta _j(t)}.
\label{order}
\end{equation}
When a synchronous state is formed, $r(t)$ takes a positive value.
Indeed, based on some formal calculations, Kuramoto assumed a bifurcation diagram of $r(t)$:
Suppose $N\to \infty$.
If $g(\omega )$ is an even and unimodal function such that $g''(0)\neq 0$, then the bifurcation
diagram of $r(t)$ is as in Fig.\ref{fig1}(a). In other words, if the coupling strength $K$ is 
smaller than $K_c := 2/(\pi g(0))$, then $r(t) \equiv 0$ is asymptotically stable.
If $K$ exceeds $K_c$, then a stable synchronous state emerges.
Near the transition point $K_c$, $r$ is of order $O((K- K_c)^{1/2})$.
See \cite{Str1} for Kuramoto's discussion.

In the last two decades, several studies have been performed in an attempt to confirm Kuramoto's assumption.
Daido~\cite{Dai} calculated steady states of Eq. (\ref{KM}) for any $f$ using an argument similar to Kuramoto's.
Although he obtained various bifurcation diagrams, the stability of solutions was not demonstrated.
In order to investigate the stability of steady states, Strogatz and Mirollo and coworker~\cite{Mir0, Mir2, Str3, Str2} performed
a linearized analysis.
The linear operator $T_1$, which is obtained by linearizing the Kuramoto model, has a continuous spectrum on the imaginary axis.
Nevertheless, they found that the steady states can be asymptotically stable because of the existence of resonance poles on the left half plane~\cite{Str2}.
Since the results of Strogatz and Mirollo and coworker are based on the linearized analysis, the effects of nonlinear terms are neglected.
However, investigating nonlinear bifurcations is more difficult 
because $T_1$ has a continuous spectrum on the imaginary axis, that is, 
a center manifold in the usual sense is of infinite dimension.
In order to avoid this difficulty, Crawford and Davies \cite{Cra3}
added a noise of strength $D>0$ to the Kuramoto model.
The continuous spectrum then moves to the left side by $D$, and thus the usual center manifold
reduction is applicable.
However, their method is not valid when $D=0$.
An eigenfunction of $T_1$ associated with a center subspace diverges as $D\to 0$
because an eigenvalue on the imaginary axis is embedded in the continuous spectrum as $D\to 0$.
Recently, Ott and Antonsen~\cite{Ott1} found a special solution of the Kuramoto model, 
which allows the dimension of the system to be reduced.
Their method is applicable only for $f(\theta ) = \sin \theta $ because their method relies on a 
certain symmetry of the system~\cite{Marv}. Furthermore, the reduced system is still of infinite dimension, 
except for the case in which $g(\omega )$ is a rational function.
Thus, a unified bifurcation theory for globally coupled phase oscillators is required.

In the present letter, a correct center manifold reduction is proposed by means of the theory of generalized functions, 
which is applicable for any coupling function $f$.
It is shown that there exists a finite-dimensional center manifold on a space of generalized functions, 
despite the fact that the continuous spectrum lies on the imaginary axis.
This will be demonstrated for two cases, (I) $f(\theta ) = \sin \theta $ and (II) $f(\theta )
 = \sin \theta + h \sin 2\theta , h\in \mathbf{R}$, and two distribution functions $g(\omega )$, 
(a) a Gaussian distribution and (b) a rational function (e.g., Lorentzian distribution $g(\omega ) = 1/(\pi (1 + \omega ^2))$).
For (I), we rigorously prove Kuramoto's conjecture, and for (II), 
a different bifurcation diagram will be obtained, as was predicted by Daido~\cite{Dai}.
The different bifurcation structure is shown to be caused by the discontinuity of the projection 
to the generalized center subspace. All omitted proofs are given in \cite{Chi}.


\textbf{Settings}.
The continuous model of Eq. (\ref{KM}) is given by $\partial \rho_t/\partial t + \partial (\rho_t v)/\partial \theta =0$
with $v := \omega  + K \sum^\infty_{l=-\infty} f_l \eta_l (t) e^{-\sqrt{-1} l \theta }$,
where $\eta_l$ is defined to be
\begin{eqnarray*}
\eta_l(t) = \int_{\mathbf{R}} \! \int^{2\pi}_{0} \!  e^{\sqrt{-1} l \theta } \rho_t (\theta , \omega ) g(\omega ) d\theta d\omega,
\end{eqnarray*}
and $\rho_t = \rho_t (\theta , \omega )$ is a probability measure on $[0, 2\pi)$ parameterized by $t, \omega \in \mathbf{R}$.
In particular, $\eta_1$ is a continuous version of Kuramoto's order parameter.
Setting $Z_j(t, \omega ) := \int^{2\pi}_{0} \! e^{\sqrt{-1}j\theta }  \rho_t (\theta , \omega ) d\theta $ yields
\begin{eqnarray}
\frac{dZ_j}{dt} &=& \sqrt{-1}j\omega Z_j +\sqrt{-1} j K f_j \eta _j + \sqrt{-1} j K \sum_{l\neq j} f_l\eta_l Z_{j-l} \nonumber \\
&:=& T_j Z_j + \sqrt{-1} j K \sum_{l\neq j} f_l\eta_l Z_{j-l}, \label{Z_j} \\
\eta _j(t) &=& \int_{\mathbf{R}} \! Z_j(t, \omega )g(\omega ) d\omega = (\overline{Z_j}, P_0) = (P_0, Z_j), \label{eta _j}
\end{eqnarray}
where $(\, \cdot\, ,\, \cdot \,)$ is the inner product on the weighted Lebesgue space $L^2 (\mathbf{R}, g(\omega )d\omega )$, and $P_0 (\omega ) \equiv 1$.
The linear operator $T_j$ is known to have a continuous spectrum on the imaginary axis.
Furthermore, there exists a positive constant $K^{(j)}_c$, such that if $K^{(j)}_c<K$, 
$T_j$ has eigenvalues on the right half plane (such that the de-synchronous state is unstable), 
while if $0<K<K^{(j)}_c$, $T_j$ has no eigenvalues.
For example, if $f$ is an odd function and if 
$g$ is even and unimodal, then $K^{(j)}_c = -\mathrm{Im}(f_j)/(\pi |f_j|^2 g(0))$.
In the present letter, for simplicity, we assume that $K_c:= \inf_j K^{(j)}_c = K^{(1)}_c$
(for $f(\theta ) = \sin \theta  + h\sin 2\theta $, which is true if and only if $h<1$).
When $0<K<K_c$, $T_j$ has no eigenvalues, and thus the dynamics of the linearized system $dZ_j/dt = T_jZ_j$ 
is quite nontrivial. In \cite{Chi}, the spectral theory on rigged Hilbert spaces is developed 
to reveal the dynamics of the linearized system.

A rigged Hilbert space consists of three spaces $X \subset L^2(\mathbf{R}, g(\omega )d\omega ) \subset X'$:
a space $X$ of test functions, a Hilbert space $L^2(\mathbf{R}, g(\omega )d\omega )$, 
and the dual space $X'$ of $X$ (a space of continuous anti-linear functionals on $X$, each element of which is referred to as a generalized function).
We use Dirac's notation, where for $\mu \in X'$ and $\phi \in X$, $\mu (\phi )$ is denoted by
$\langle \phi \,|\, \mu \rangle$. For $a \in \mathbf{C}$, 
we have $a \langle \phi \,|\, \mu \rangle = \langle \overline{a} \phi \,|\, \mu \rangle = \langle \phi \,|\, a \mu \rangle$.
The canonical inclusion $i: X \to X'$ is defined as follows. For $\psi\in X$,
we denote $i(\psi)$ by $| \psi \rangle$, which is defined to be
\begin{equation}
i(\psi)(\phi) = \langle \phi \,|\, \psi \rangle := (\phi, \psi)
 = \int_{\mathbf{R}}\! \overline{\phi (\omega )}\psi (\omega )g(\omega )d\omega .
\end{equation}
By the canonical inclusion, Eq. (\ref{Z_j}) is rewritten as an evolution equation on the dual space $X'$ as
\begin{eqnarray}
\frac{d}{dt} |\, Z_j \rangle
  = T_j^\times |\, Z_j \rangle 
     + \sqrt{-1} j K \sum_{l\neq j} f_l \langle P_0 \,|\, Z_l \rangle 
           \cdot |\, Z_{j-l} \rangle, \label{dual}
\end{eqnarray}
where $T_j^\times$ is a dual operator of $T_j$ defined through
$\langle \phi \,|\, T_j^\times\mu \rangle = \langle T_j \phi \,|\, \mu \rangle$.

Here, the strategy for the bifurcation theory of globally coupled phase oscillators is 
to use the space of generalized functions $X'$ rather than a space of usual functions.
The reason for this is explained intuitively as follows.
If we use the space $L^2(\mathbf{R}, g(\omega )d\omega )$ to investigate the dynamics,
then the behavior of $\rho_t$ itself will be obtained. 
However, it is neutrally stable because of the conservation law $\int^{2\pi}_{0}\! \rho_t (\theta , \omega )d\theta =1$.
What we would like to know is the dynamics of the moments of $\rho_t$, in particular, the order parameter. 
This suggests that we should use a different topology for the stability of $\rho_t$. 
(Note that the definition of stability depends on definition of the topology.) 
For the purpose of the present study, $\rho_t$ is said to be convergent to $\hat{\rho}$ as $t\to \infty$ if and only if 
\begin{eqnarray*}
\int_{\mathbf{R}}\! \int^{2\pi}_{0}\! \phi(\omega ) e^{\sqrt{-1}j\theta } d\rho_t (\theta , \omega ) \to 
\int_{\mathbf{R}}\! \int^{2\pi}_{0}\! \phi(\omega ) e^{\sqrt{-1}j\theta }d\hat{\rho} (\theta , \omega ) 
\end{eqnarray*}
for any $j\in \mathbf{Z}$ and $\phi \in X$.
The topology induced by this convergence is referred to as the weak topology.
By the completion of $L^2(\mathbf{R}, g(\omega )d\omega )$ with respect to the weak topology,
we obtain a space of generalized functions $X'$.
On the space $X'$, a function $Z_1(t, \omega )$ converges as $t\to \infty$ if and only if  
$\langle \phi \,|\, Z_1 \rangle$ converges for any $\phi \in X$.
Since the order parameter is written as $\eta_1 (t) = \langle P_0 \,|\, Z_1 \rangle$,
this topology is suitable for the purpose of the present study.
For this topology, it turns out that $\rho_t$ is not neutrally stable.

A suitable choice of $X$ depends on $g(\omega )$. 
When $g(\omega )$ is a Gaussian distribution, let $\mathrm{Exp}_+(\beta)$ be the set of holomorphic functions 
defined near the upper half plane such that $|\phi (z)|e^{-\beta |z|}$ is finite. 
Set $X = \mathrm{Exp}_+ := \bigcup_{\beta \geq 0}\mathrm{Exp}_+(\beta)$. 
We can introduce a suitable topology on $\mathrm{Exp}_+$ so that the dual space $\mathrm{Exp}_+'$ becomes 
a complete metric space, which allows the existence of a center manifold on $\mathrm{Exp}_+'$ to be proven.
When $g(\omega )$ is a rational function,
$X := H_+$ is a space of bounded holomorphic functions on the real axis and the upper half plane.
In this case, we can show that $i(H_+) \subset H_+'$ is a finite-dimensional vector space, 
which implies that Eq. (\ref{dual}) is essentially a finite-dimensional system.
This is why in \cite{Ott1, Mart}, the system is reduced to a finite-dimension system
when $g(\omega )$ is the (sum of the) Lorentzian distribution.

Let $e^{T_1t}$ be the semigroup generated by $T_1$. In \cite{Chi}, the spectral decomposition of $e^{T_1t}$ is obtained by means of the rigged Hilbert space.
Define resonance poles $\lambda _0, \lambda _1,\cdots $ of $T_1$ to be roots of the equation
\begin{equation}
\int_{\mathbf{R}}\! \frac{1}{\lambda - \sqrt{-1}\omega }g(\omega )d\omega 
  + 2\pi g (-\sqrt{-1} \lambda ) = \frac{2}{K},
\label{resonance}
\end{equation}
on the imaginary axis and the left half plane.
Roughly speaking, a resonance pole is a continuation of an eigenvalue to the second Riemann sheet
of the resolvent $(\lambda - T_1)^{-1}$~\cite{Str2, Chi}.
In the following, for simplicity, we assume that all $\lambda _n$'s are single roots of Eq. (\ref{resonance}). Define a functional $\mu_j \in X'$ to be
\begin{eqnarray*}
\langle \phi \,|\, \mu_j  \rangle = 
\left\{ \begin{array}{ll}
\displaystyle \int_{\mathbf{R}} \! \frac{\overline{\phi (\omega )}g(\omega )}{\lambda_j -\sqrt{-1}\omega }d\omega & \\
\quad +  2\pi \overline{\phi (-\sqrt{-1}\lambda_j )}g(-\sqrt{-1}\lambda_j ), & (\mathrm{Re}(\lambda_j ) < 0), \\[0.3cm]
\displaystyle \lim_{x \to +0} \!
\int_{\mathbf{R}} \! \frac{\overline{\phi (\omega )}g(\omega )}{(x\!+\! \sqrt{-1}y_j) -\sqrt{-1}\omega }d\omega, & 
(\lambda_j = \sqrt{-1}y_j). \\
\end{array} \right.
\end{eqnarray*}
The $\mu _j$'s are called the generalized eigenfunctions associated with the resonance poles 
due to the equality $T_1^\times |\, \mu_j \rangle = \lambda _j |\, \mu_j \rangle$.
Then, we can prove that the semigroup is expressed as
\begin{equation}
(e^{T_1t})^\times | \psi \rangle = \sum^\infty_{n=0} \frac{K}{2D_n}e^{\lambda _nt}
\langle \psi  \,|\, \mu_n \rangle \cdot | \mu_n \rangle,
\label{decom}
\end{equation}
for any $\psi \in X$, which gives the spectral decomposition of $e^{T_1t}$ on the dual space $X'$,
where $D_n$ are constants defined by
\begin{eqnarray*}
D_n = \lim_{\lambda \to \lambda _n} \frac{1}{\lambda - \lambda _n}
\left( 1 - \frac{K}{2}\int_{\mathbf{R}} \! \frac{g(\omega )}{\lambda -\sqrt{-1}\omega }d\omega 
           - \pi K g(-\sqrt{-1}\lambda )\right).
\end{eqnarray*}
(In the previous paragraph, $X$ was chosen so that the right-hand side of Eq.(\ref{decom}) converges.)
Equation (\ref{decom}) completely determines the dynamics of the order parameter for the linearized system. 
If $0<K<K_c$, then all resonance poles $\lambda _n$ lie on the left half plane.
As a result, $\eta_1(t) = (e^{T_1t}\phi, P_0) = \langle \phi \,|\, (e^{T_1t})^\times P_0 \rangle$ 
decays to zero exponentially as $t\to \infty$, which proves the asymptotic stability
of the de-synchronous state.


\textbf{Center manifold reduction}.
When $K=K_c$, there exist resonance poles on the imaginary axis.
The generalized center subspace $\mathbf{E}_c \subset X'$ is defined as a space spanned by
generalized eigenfunctions associated with resonance poles on the imaginary axis,
say $\lambda _0,\cdots , \lambda _M$.
Equation (\ref{decom}) suggests that the projection $\Pi_c$ to $\mathbf{E}_c$ is given by
\begin{equation}
\Pi_c | \psi \rangle = \sum^M_{n=0} \frac{K}{2D_n}
\langle  \psi \,|\, \mu_n \rangle \cdot |\, \mu_n \rangle .
\end{equation}
In general, $\Pi_c : X' \to X'$ is continuous only on a subspace of $X'$ because the topology on $X'$ is too weak. 
When $X = \mathrm{Exp}_+$, it is proven in \cite{Chi} that $\Pi_c$ is continuous only on $i(\mathrm{Exp}_+(0))$.
For a solution of Eq. (\ref{Z_j}), $Z_1, Z_2, \cdots $ are included in $\mathrm{Exp}_+(0)$,
although $Z_{-1}, Z_{-2},\cdots $ are not. Because of the discontinuity of $\Pi_c$, 
an interesting bifurcation occurs when $f(\theta ) \neq \sin \theta $.
In what follows, assume that $g$ is an even and unimodal function. In this case, 
on the imaginary axis, there is only one resonance pole $\lambda _0 = 0$ at $K=K_c$. 
Hence, $\mathbf{E}_c = \mathrm{span} \{ \mu_0 \}$ is of one dimension, where $\mu_0$ is the 
generalized eigenfunction associated with $\lambda _0 = 0$.
Next, let us derive the dynamics on a center manifold.
The derivation is performed in the same way for both (a) a Gaussian distribution and 
(b) rational functions.

\textbf{(I)} First, we consider $f(\theta ) = \sin \theta $.
In this case, equations for $Z_1, Z_2, \cdots $ do not depend on $Z_{-1}, Z_{-2},\cdots $.
Thus, $\Pi_c$ acts continuously on solutions of Eq. (\ref{dual}).
Set $\varepsilon = K-K_c$. Then, Eq. (\ref{dual}) for $j=1$ is rewritten as
\begin{equation}
\frac{d}{dt} |\, Z_1 \rangle
  = T_{10}^\times |\, Z_1 \rangle  
       + \frac{\varepsilon }{2} \langle  P_0 \,|\, Z_1 \rangle |\, P_0  \rangle
           - \frac{K}{2} \langle P_0 \,|\, Z_1 \rangle |\, Z_2 \rangle,
\label{Z_1}
\end{equation}
where $T_{10}$ is defined by replacing $K$ with $K_c$ in the definition of $T_1$.
In order to obtain the dynamics on the center manifold, using the spectral decomposition, we set
\begin{equation}
|\, Z_1 \rangle = \frac{K_c}{2}\alpha |\, \mu_0 \rangle + |\, Y_1 \rangle,
\,\, \alpha (t) = \frac{1}{D_0} \langle Z_1 \,|\, \mu_0 \rangle,
\end{equation}
along the direct sum $\mathbf{E}_c \oplus \mathbf{E}_c^{\bot}$.
The purpose here is to derive the dynamics of $\alpha $.
Since $|\, Y_1 \rangle$ and $|\, Z_2 \rangle$ are included in the stable subspace,
the center manifold theorem~\cite{Chi} reveals that on the center manifold, 
$|\, Y_1 \rangle,\, |\, Z_2 \rangle \sim O(\alpha ^2)$.
In particular, the last term $\langle P_0 \,|\, Z_1 \rangle |\, Z_2 \rangle$ 
of Eq. (\ref{Z_1}) is of order $O(\alpha ^3)$.
Apply the projection $\Pi_c$ to both sides of Eq. (\ref{Z_1}).
Noting that $\langle P_0 \,|\, Z_1 \rangle  \Pi_c |\, Z_2 \rangle$ is also of 
order $O(\alpha ^3)$ because the projection is continuous on $|\, Z_2 \rangle$, 
we obtain the dynamics on the center manifold as
\begin{equation}
\frac{d}{dt}\alpha  = \frac{\alpha }{D_0 K_c} \left( 
   \varepsilon  + \frac{\pi g''(0)K_c^4}{16} |\alpha |^2 \right) 
         + O(\varepsilon \alpha ^2, \varepsilon ^2\alpha , \varepsilon ^3, \alpha ^4).
\label{center}
\end{equation}
If $g''(0) <0$, this equation has a fixed point of order $O((K-K_c)^{1/2})$
when $\varepsilon  = K- K_c >0$. 
It is easy to verify that the order parameter $\eta _1(t)$ is given by
$\eta_1 (t) = \alpha + \mathrm{h.o.t.}$
Thus, the dynamics of the order parameter is also given by Eq. (\ref{center}).
Since $D_0 >0$, when $g$ is even and unimodal, $\alpha =0$ (de-synchronous state) is unstable, 
and the nontrivial fixed point (synchronous state) is asymptotically stable when $\varepsilon  = K- K_c >0$, 
which confirms Kuramoto's diagram.

\textbf{(II)} Assume that $f(\theta ) = \sin \theta  + h\sin 2\theta $ with $h\in \mathbf{R}$.
Then, Eq. (\ref{dual}) for $j=1$ is given by
\begin{eqnarray}
& & \frac{d}{dt} |\, Z_1 \rangle
  = T_{10}^\times |\, Z_1 \rangle  
       + \frac{\varepsilon }{2} \langle  P_0 \,|\, Z_1 \rangle |\, P_0  \rangle \nonumber \\
& - & \frac{K}{2} \left( \langle P_0 \,|\, Z_1 \rangle |\, Z_2 \rangle
    +                  h \langle P_0 \,|\, Z_2 \rangle |\, Z_3 \rangle
    -                  h \langle P_0 \,|\, Z_2 \rangle |\, Z_{-1} \rangle\right). \quad
\label{Z_1b}
\end{eqnarray}
In this case, $Z_{-1}$, on which $\Pi_c$ is discontinuous, appears.
As before, $|\, Z_2 \rangle$ satisfies $|\, Z_2 \rangle \sim O(\alpha ^2)$
and $\Pi_c|\, Z_2 \rangle \sim O(\alpha ^2)$ since $\Pi_c$ acts continuously on $|\, Z_2 \rangle$.
This implies that the term $\langle P_0 \,|\, Z_1 \rangle |\, Z_2 \rangle$ in Eq. (\ref{Z_1b})
yields a cubic nonlinearity as case (I).
On the other hand, since $\Pi_c$ is discontinuous on $Z_{-1}$, we can show that 
$\Pi_c |\, Z_{-1} \rangle \sim O(1)$ even if $|\, Z_{-1} \rangle \sim O(\alpha )$.
As a result, the last term $\langle P_0 \,|\, Z_2 \rangle |\, Z_{-1} \rangle$
in Eq. (\ref{Z_1b}) yields a quadratic nonlinearity.
Indeed, we can verify that
\begin{equation}
\Pi_0 |\, Z_{-1} \rangle = \frac{\pi K_c g(0)}{D_0} e^{- \sqrt{-1}\mathrm{arg}(\alpha )} + O(\alpha ).
\label{P0Z1}
\end{equation}
Applying the projection $\Pi_c$ to both sides of Eq. (\ref{Z_1b}), we obtain the dynamics on the center manifold as
\begin{equation}
\frac{d \alpha }{dt} = \frac{\alpha }{D_0 K_c}
    \left( \varepsilon  - \frac{K_c^3 Ch}{2(1-h)} \alpha e^{-\sqrt{-1}\mathrm{arg}(\alpha )}\right) + \mathrm{h.o.t.},
\label{trans}
\end{equation}
where $C := \mathrm{p.v.} \int_{\mathbf{R}}\! g'(\omega )/\omega d\omega$ is a negative constant,
which proves that there exists a fixed point that is expressed as
\begin{equation}
|\eta_ 1| \sim |\alpha | = \frac{2(1-h)}{K_c^3 C h}(K-K_c) + O((K-K_c)^2).
\label{trans2}
\end{equation}
Therefore, for $h<0$, a stable branch emerges when $K_c < K$, and for $0<h<1$, an unstable branch emerges when $K < K_c$ (Fig.\ref{fig1}(b)). 

\begin{figure}
\includegraphics{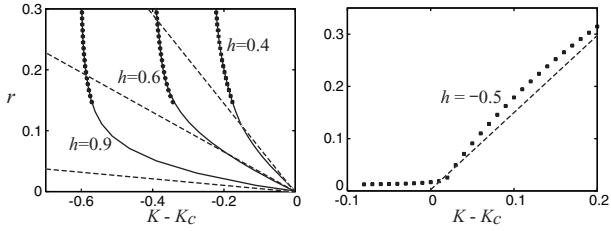}
\caption{Numerical results for $f(\theta ) = \sin \theta + h\sin 2 \theta $.
Black dots denote the order parameter calculated from Eq. (\ref{KM}) 
for $N=8,000, g(\omega ) = e^{-\omega ^2/2}/\sqrt{2\pi}$ using the method shown in \cite{Ton}. 
Since $r$ is unstable when $K-K_c < 0$, it is difficult to obtain small $r$.
The solid lines are interpolations of black dots using quintic polynomials.
The dotted lines denote the analytical results obtained by Eq. (\ref{trans2}).
\label{fig2}}
\end{figure}


\textbf{Discussion.}
Equation (\ref{dual}) shows that the dynamics of $Z_1, Z_2,\cdots $ is independent of 
$Z_{-1}, Z_{-2},\cdots $ if and only if $f(\theta ) = \sin \theta $. 
In other words, Eq. (\ref{dual}) splits into two systems: a system of $\{ Z_1, Z_2, \cdots \}$
and a system of $\{ Z_{-1}, Z_{-2}, \cdots \}$.
Since the projection $\Pi_c$ is continuous on a solution of the former system,
we can show the existence of a smooth center manifold.
Note that Eq. (\ref{KM}) is invariant under the rotation on a circle.
As a result, the dynamics on the center manifold is also invariant under the rotation $\alpha \mapsto e^{\sqrt{-1}\phi}\alpha $.
If a center manifold is smooth, then the dynamics on this manifold with the rotation symmetry
must be of the form $\dot{\alpha } = \alpha F(|\alpha |^2)$.
Thus, a cubic nonlinearity is dominant, and a pitchfork bifurcation generally occurs, as shown in Eq. (\ref{center}).
On the other hand, if $f(\theta ) \neq \sin \theta $, 
then the equations of $Z_{1}, Z_{2},\cdots $ depend on $Z_{-1}, Z_{-2},\cdots $, on which $\Pi_c$ is not continuous.
In such a case, the center manifold is not smooth, and quadratic nonlinearity may appear, as described above. 
In this manner, different bifurcations occur when $f(\theta ) \neq \sin \theta $.
Although the diagram shown in Fig.\ref{fig1}(b) looks like a transcritical bifurcation,
Eq. (\ref{trans}) is different from the normal form of a transcritical bifurcation.
Because of the factor $e^{- \sqrt{-1}\mathrm{arg}(\alpha )}$ caused by the discontinuity of $\Pi_c$, 
Eq. (\ref{trans}) remains invariant under the rotation despite the existence of a quadratic nonlinearity. 
The discontinuity induces a new type of bifurcation including $e^{- \sqrt{-1}\mathrm{arg}(\alpha )}$.

A center manifold reduction for globally coupled phase oscillators was also developed 
by Crawford and Davies \cite{Cra3} with a noise of strength $D>0$.
Although they also expected a diagram such as shown as Fig.\ref{fig1}(b) when $D=0$, 
the factor $e^{- \sqrt{-1}\mathrm{arg}(\alpha )}$ was not obtained. Since the eigenfunction diverges as $D\to 0$, 
expressions of the dynamics on the center manifold were not shown explicitly.
In the present letter, we have shown that the eigenfunction $\mu_0$ exists on a space of generalized functions, 
which provides a correct center manifold reduction.
The diagram shown in Fig.\ref{fig1}(b) was also obtained by Daido~\cite{Dai} by means of a self-consistent analysis. 
Unfortunately, his results were not correct because he performed inappropriate termwise integrations of certain infinite series.
According to his results, the order parameter is given as $(1-2h) \cdot \mathrm{const}.$, 
which suggests that some degeneracy occurs when $h=1/2$.
However, the numerical results given in Fig.\ref{fig2} show that the critical exponent of the 
order parameter changes only when $h=1$, which agrees with the results of the present study (\ref{trans2}). 
Ott and Antonsen~\cite{Ott1} found an inertia manifold given by $Z_n = (Z_1)^n$ when $f(\theta ) = \sin \theta $. 
The center manifold of the present study is a finite-dimensional submanifold of the inertia manifold, 
which provides a further reduction of the results of Ott and Antonsen.
The key strategy of the present theory is to use spaces of generalized functions and the weak topology. 
The weak topology is suitable for investigating the dynamics of moments of probability density functions. 
Since the strategy is independent of the details of the models, this strategy will
be extended to various types of large populations of coupled systems and evolution equations of density functions, 
such as the Vlasov equation.
 
\begin{acknowledgments}
The present study was supported by Grant-in-Aid for Young Scientists (B), No.22740069 from MEXT Japan.
\end{acknowledgments}


%

\end{document}